\documentclass{aa}
\usepackage{graphicx}
\usepackage[varg]{txfonts}
\usepackage{natbib}
\usepackage{color}

\begin{document}

\title{Significantly high polarization degree of the very low-albedo asteroid (152679) 1998 KU$_\mathrm{2}$}

\subtitle{} 

\author{Daisuke Kuroda\inst{1}
\and  Masateru Ishiguro\inst{2} 
\and  Makoto Watanabe\inst{3}
\and Sunao Hasegawa\inst{4}
\and Tomohiko Sekiguchi\inst{5}
\and Hiroyuki Naito\inst{6}
\and Fumihiko Usui\inst{7}
\and Masataka Imai\inst{8}
\and Mitsuteru Sato\inst{8}
\and Kiyoshi Kuramoto\inst{8}
}

\institute{
Okayama Astrophysical Observatory, National Astronomical Observatory of Japan, Asakuchi, Okayama 719-0232, Japan
  \and Department of Physics and Astronomy, Seoul National University, Gwanak, Seoul 151-742, Korea
  \and Department of Applied Physics, Okayama University of Science,  Ridai-cho, Kita-ku, Okayama, Okayama 700-0005, Japan
  \and Institute of Space and Astronautical Science, Japan Aerospace Exploration Agency, Sagamihara, Kanagawa 252-5210, Japan
  \and Asahikawa Campus, Hokkaido University of Education, Hokumon, Asahikawa, Hokkaido 070-8621, Japan
  \and Nayoro Observatory, Nisshin, Nayoro, Hokkaido 096-0066, Japan 
  \and Center for Planetary Science, Graduate School of Science, Kobe University, Minatojima-Minamimachi, Chuo-Ku, Kobe, Hyogo 650-0047, Japan
    \and Department of Cosmosciences, Graduate School of Science, Hokkaido University, Kita-ku, Sapporo, Hokkaido 060-0810, Japan
 } 

\date{Received 11 October 2017 / Accepted 15 November 2017}

\bibpunct{(}{)}{;}{a}{}{,} 

\abstract {
We present a unique and significant polarimetric result regarding the near-Earth asteroid (152679) 1998 KU$_\mathrm{2}$ , which
has a very low geometric albedo. From our observations, we find that the linear polarization degrees of 1998 KU$_\mathrm{2}$ are 44.6 $\pm$ 0.5\% in the R$_\mathrm{C}$ band and 44.0 $\pm$ 0.6\% in the V band at a solar phase angle of 81.0\degr. These values are the highest of any known airless body in the solar system (i.e., high-polarization comets, asteroids, and planetary satellites) at similar phase angles. This polarimetric observation is not only the first for primitive asteroids at large phase angles, but also for low-albedo (< 0.1) airless bodies.

Based on spectroscopic similarities and polarimetric measurements of materials that have been sorted by size in previous studies, we conjecture that 1998 KU$_\mathrm{2}$ has a highly microporous regolith structure comprising nano-sized carbon grains on the surface.
}


\maketitle
%
\section{Introduction}
The polarimetric research of asteroids has attracted attention as a suitable method for investigating their surface properties. The sunlight scattered on an asteroidal surface can be measured as a partially linearly polarized quantity because it is affected by some scattering features of the surface layer (e.g., composition, albedo, roughness, and structure) and the solar phase angle $\alpha$ (the angle between the Sun and the observer as seen from the asteroid). 

Previous polarimetric studies have focused on polarization as a function of $\alpha$, which is called the polarization phase curve \citep{Muinonen:2002}. In such a curve, there are two major trends, comprising a negative-polarization branch in the region of $\alpha\lesssim$ 20\degr \ and a positive-polarization branch including its maximum value around $\alpha\sim$ 90\degr--100\degr. The signals corresponding to the negative branch, which are mainly collected for main-belt asteroids, have a prominent polarization component parallel to the scattering plane and are attributable to coherent backscattering of sunlight \citep{Shkuratov:1985, Muinonen:2002}. On the other hand, the signals in the positive branch, which are acquired with observations of near-Earth asteroids (hereafter NEAs), are dominated by the perpendicular component of the phase angle to the peak. Such polarimetric behaviors have been estimated in terms of correlations with the geometric albedo or asteroid taxonomic type based on statistical research as summarized in \citet{Belskaya:2015}.

Very few asteroids corresponding to the positive branch have been studied in detail. In this regard, six of the taxonomic S-complex\footnote{S-type asteroids in a broad sense, including S subgroups.} NEAs with moderate albedo and only one E-type NEA with high albedo were observed at phase angles larger than 80\degr. It is reported that these asteroids present polarization degrees smaller than 10\% \citep{Kiselev:1990,Kiselev:2002, Ishiguro:1997,Ishiguro:2017, Delbo:2007, Belskaya:2009b, Fornasier:2015}. Recently, NEA (3200) Phaethon with an intermediate albedo \citep[$p_\mathrm{V}$ = 0.12,][]{Hanus:2016} showed a polarization degree of up to 50\%; the authors presumed that a paucity of small grains in the asteroid surface boosted the polarization degree \citep{Ito:2017}. Little is known about the polarimetric properties of dark asteroids (i.e., $p_\mathrm{V}$ < 0.1), such as the taxonomic C-complex\footnote{C-type asteroids in a broad sense, including C subgroups.}, at large phase angles ($\alpha$ > 40\degr).

For solar system objects with a geometric albedo lower than 0.1, polarimetric observations at large phase angles have been conducted for many comets and the two satellites of Mars. Statistical studies of these comets have revealed a maximum polarization degree ($P_\mathrm{max}$) of 25\%--28\% at $\alpha\sim$ 90\degr--100\degr \ on the positive branch. Thus, the dark asteroids are expected to display similar polarimetric behaviors because their albedos are similar. However, it may be difficult to directly compare asteroids and Martian satellites with comets because some cometary components (i.e., not only the nuclei, but also gas, dust, jet, etc.) affect the polarization. The approaches for deriving the polarization of the cometary nuclei (e.g., airless bodies) of 2P/Encke and 209P/LINEAR have yielded polarization degrees of $P\sim$ 30\%--40\% around $\alpha\sim$ 90\degr--100\degr \citep{Jockers:2005, Kuroda:2015}.

In this work, we measured the linear polarization degree for a very dark NEA, (152679) 1998 KU$_\mathrm{2}$, at multiple large phase angles. 1998 KU$_\mathrm{2}$ has a geometric albedo of 0.018--0.03 \citep{Mainzer:2011, Nugent:2016}, which is significantly lower than those of the majority of asteroids \citep[typically,
these are 0.26 for S-type and 0.08 for C-type asteroids,][]{DeMeo:2013} even though we took the error into account (0.006 or less). This NEA is classified as a taxonomic F-type \citep{Whiteley:2001} or Cb-type \citep{Binzel:2004} asteroid; these asteroid types are considered as primitive bodies containing organic materials. Our observations reveal that this very low-albedo NEA has uncommon polarimetric features. 1998 KU$_\mathrm{2}$ exhibits a significantly high polarization degree, even higher than that of Phaethon at the same phase angle. Our findings may aid in not only extending the relationship between the geometric albedo and the polarization degree, but also in demonstrating the possible existence of asteroids
with very low albedo and/or suggest new approaches to understanding primitive materials in the solar system.
\section{Observations and reductions}
\subsection{Observations}
Optical polarimetric measurements of 1998 KU$_\mathrm{2}$ were carried out for three nights in June-July, 2015, using the visible Multi-Spectral Imager (hereafter MSI) on the 1.6 m Pirka telescope at Hokkaido University's Nayoro Observatory in Hokkaido, Japan. To the MSI, an EM-CCD camera (Hamamatsu Photonics C9100-13) with a back-thinned frame transfer CCD of 512 $\times$ 512 pixels (pixel scale of 16 $\mu$m, 0.389 $\arcsec$/pixel) and the Johnson-Cousins filter system \citep{Watanabe:2012} is attached. The imaging polarimetric mode of the MSI, which implements a rotatable half-wave plate as the polarimetric modulator and a Wollaston prism as the beam
splitter, simultaneously generates ordinary and extraordinary images with perpendicular polarizations at four position angles ($\theta$ = 0\degr, 45\degr, 22.5\degr, and 67.5\degr).  The use of this method allows for the cancellation of errors caused by any atmospheric fluctuations.

We observed 1998 KU$_\mathrm{2}$ through an R$_\mathrm{C}$-band filter at three phase angles, $\alpha$ = 49.8\degr, 50.7\degr, and 81.0\degr,  and a V filter at only $\alpha$ = 81.0\degr. The target was tracked with non-sidereal motion of the telescope, and the exposure times were chosen as 60 or 90 seconds depending on its brightness. In each frame, 1998 KU$_\mathrm{2}$ was captured as point-source images (see section 3.3). The observation circumstances are summarized in Table~\ref{table1}.
\begin{table*}
\caption{\label{table1}Observation circumstances of (152679) 1998 KU$_\mathrm{2}$}
\centering
\begin{tabular}{lcccccccc}
\hline\hline
Date & UT & Filter\tablefootmark{a} & Exp.\tablefootmark{b} & Vmag\tablefootmark{c}  & r\tablefootmark{d} & $\Delta\tablefootmark{e}$ & $\alpha\tablefootmark{f}$ & $\phi\tablefootmark{g}$ \\
 & & & [s] & [V] & [au] & [au] & [\degr] & [\degr] \\
\hline
2015-Jun-11 & 15:44:47-15:49:07 & R$_\mathrm{C}$ & 60 & 16.12 & 1.17813 & 0.29000 & 49.804 & 241.85 \\ 
2015-Jun-12 & 16:40:25-16:44:45 & R$_\mathrm{C}$ & 60 & 16.10 & 1.17176 & 0.28511 & 50.721 & 241.52 \\  
2015-Jun-12 & 16:45:00-16:49:20 & R$_\mathrm{C}$ & 60 & 16.10 & 1.17175 & 0.28509 & 50.724 & 241.51 \\ 
2015-Jun-12 & 16:49:31-16:53:51 & R$_\mathrm{C}$ & 60 & 16.10 & 1.17173 & 0.28508 & 50.727 & 241.51 \\ 
2015-Jul-16 & 16:23:58-16:30:21 & R$_\mathrm{C}$ & 90 & 16.72 & 1.02169 & 0.28037 & 81.040 & 249.38 \\ 
2015-Jul-16 & 16:30:43-16:37:07 & R$_\mathrm{C}$ & 90 & 16.72 & 1.02168 & 0.28038 & 81.041 & 249.38 \\ 
2015-Jul-16 & 16:37:18-16:43:42 & R$_\mathrm{C}$ & 90 & 16.72 & 1.02167 & 0.28040 & 81.042 & 249.38 \\
2015-Jul-16 & 16:43:53-16:50:18 & R$_\mathrm{C}$ & 90 & 16.72 & 1.02166 & 0.28041 & 81.044 & 249.39 \\
2015-Jul-16 & 16:50:34-16:56:57 & V & 90 & 16.72 & 1.02165 & 0.28043 & 81.046 & 249.39 \\ 
2015-Jul-16 & 16:57:09-17:03:33 & V & 90 & 16.72 & 1.02164 & 0.28044 & 81.047 & 249.39  \\
2015-Jul-16 & 17:03:44-17:10:08 & V & 90 & 16.72 & 1.02163 & 0.28046 & 81.049 & 249.39 \\ 
2015-Jul-16 & 17:11:00-17:17:25 & R$_\mathrm{C}$ & 90 & 16.72 & 1.02162 & 0.28047 & 81.050 & 249.39 \\ 
2015-Jul-16 & 17:17:36-17:23:58 & R$_\mathrm{C}$ & 90 & 16.72 & 1.02161 & 0.28049 & 81.052 & 249.40 \\  
2015-Jul-16 & 17:24:10-17:30:33 & R$_\mathrm{C}$ & 90 & 16.72 & 1.02160 & 0.28050 & 81.053 & 249.40 \\
\hline
\hline
\end{tabular}
\tablefoot{
\tablefoottext{a}{Employed filters (V: center 545, width 87 nm R$_\mathrm{C}$: center 641, width 149 nm).}
\tablefoottext{b}{Typical exposure time per frame.}
\tablefoottext{c}{Apparent visual magnitude.}
\tablefoottext{d}{Heliocentric distance.}
\tablefoottext{e}{Geocentric distance.}
\tablefoottext{f}{Solar phase angle.}
\tablefoottext{g}{Position angle of the scattering plane (in degrees E of N).}
}
\end{table*}
\subsection{Data reductions}
The obtained data of 1998 KU$_\mathrm{2}$ were processed through an original analysis pipeline (MSIRED), which handled bias subtraction, flat fielding by dome, and cosmic-ray rejection with adeptness using standard tasks within the IRAF reduction package \citep{Tody:1993}. The ordinary and extraordinary intensities were measured by circular aperture photometry using the IRAF apphot task. This method enables extracting the intensity by subtracting the surrounding sky background from the total of the pixel counts within the desired circular aperture. We evaluated the quantities with an aperture radius of 1.5 times the full width at half-maximum (FWHM) of the Moffatt point spread function \citep{Moffat:1969}. 
The normalized Stokes parameters \citep{Bohren:1983, Tinbergen:1996}, Q/I and U/I, are derived from the ordinary (o) and extraordinary (e) intensities at a given set of half-wave plate position angles in degree, which are given as

\begin{equation}
\frac{Q}{I} = \left(\sqrt{\frac{I_{e,0} / I_{o,0}}{ I_{e,45} / I_{o,45}}}-1\right) / \left(\sqrt{\frac{ I_{e,0} / I_{o,0}}{I_{e,45} / I_{o,45}}}+1\right),
 \label{eq:Q/I}
\end{equation}
\noindent
and
\begin{equation}
\frac{U}{I} = \left(\sqrt{\frac{ I_{e,22.5} / I_{o,22.5}}{ I_{e,67.5} / I_{o,67.5} }}-1\right) / \left(\sqrt{\frac{ I_{e,22.5} / I_{o,22.5} }{ I_{e,67.5} / I_{o,67.5}}}+1\right).
\label{eq:U/I}
\end{equation}
\noindent
The degree of linear polarization ($P$) and the position angle of polarization ($\theta_P$) are computed as 
\begin{equation}
P = \sqrt{ \left(\frac{Q}{I}\right)^{2} + \left(\frac{U}{I}\right)^{2} },
\label{eq:P}
\end{equation}
\noindent
and
\begin{equation}
\theta_P = \frac{1}{2} \tan^{-1} \left(\frac{U}{Q}\right),
\label{eq:theta}
\end{equation}
\noindent
respectively. 
In the process of Eqs. (1)--(4),  we corrected for the polarization efficiency, instrument polarization, and position angle zero-point using the results for polarized standard stars (HD204827, HD154445, and HD155197) and unpolarized standard stars (HD212311 and BD +32 3739). These correction terms were determined separately \citep[see Appendix, ][]{Ishiguro:2017}, which were referred to as the polarization degrees and position angles listed in \citet{Schmidt:1992}. The calibration data were taken about one month before our observation,  but a slight change in the polarimetric performance due to degradation has insignificant influence in our results. 

As a common approach to quantify the polarization for solar system objects, we represent the degree of linear polarization ($P_\mathrm{r}$) and the position angle of polarization ($\theta_\mathrm{r}$) referenced to the scattering plane \citep{Zellner:1976} as
\begin{equation}
P_r = P\cos(2\theta_r), 
 \label{eq:Pr}
\end{equation}
\noindent
and
\begin{equation}
\theta_r = \theta_P - (\phi \pm 90\degr),
 \label{eq:thetar}
\end{equation}
\noindent
where $\phi$ represents the position angle of the scattering plane (see Table~\ref{table1}), and the sign inside the bracket is chosen to satisfy 0\degr~\lid~($\phi~\pm$ 90\degr) \lid~180\degr \citep{Chernova:1993}. The results of the polarization degrees ($P$ and $P_\mathrm{r}$) and the position angles ($\theta_P$ and $\theta_\mathrm{r}$)  in each night are listed in Table~\ref{table2}. The errors of these quantities were derived from flux errors and uncertainties with each correction term through the law of propagation of errors. 
For more details on the error estimations, we refer to \citet{Ishiguro:2017}.

\begin{table*}
\caption{\label{table2}Measurement results of our polarimetric observations}
\centering
\begin{tabular}{lcccccc}
\hline\hline
Date & Filter\tablefootmark{a} & $\alpha\tablefootmark{b}$ & $P\pm\sigma_P$\tablefootmark{c} & $\theta_P\pm\sigma_{\theta_P}$\tablefootmark{d} & $P_\mathrm{r}\pm\sigma_{P_\mathrm{r}}$\tablefootmark{e} & $\theta_\mathrm{r}\pm\sigma_{\theta_\mathrm{r}}$\tablefootmark{f} \\
& & [\degr] &  [\%] & [\degr] & [\%] & [\degr] \\
\hline
2015-Jun-11 & R$_\mathrm{C}$ & 49.804 & 17.14$\pm$0.39 & 150.72$\pm$0.84 & 17.13$\pm$0.39 & -1.12$\pm$0.84 \\ 
\hline
2015-Jun-12 & R$_\mathrm{C}$ & 50.721 & 18.69$\pm$0.48 & 149.81$\pm$0.88 & 18.66$\pm$0.48 & -1.70$\pm$0.87 \\ 
2015-Jun-12 & R$_\mathrm{C}$ & 50.724 & 18.70$\pm$0.54 & 147.11$\pm$1.11 & 18.48$\pm$0.55 & -4.41$\pm$1.11 \\ 
2015-Jun-12 & R$_\mathrm{C}$ & 50.727 & 18.79$\pm$0.87 & 150.89$\pm$1.56 & 18.79$\pm$0.87 & -0.62$\pm$1.56 \\ 
 \multicolumn{3}{c} {Weighted Mean}  & 18.71$\pm$0.33 & 149.11$\pm$0.62 & 18.61$\pm$0.33 & -2.42$\pm$0.62 \\ 
 \hline
2015-Jul-16 & R$_\mathrm{C}$ & 81.040 & 44.75$\pm$1.12 & 157.05$\pm$0.73 & 44.60$\pm$1.12 & -2.33$\pm$0.73 \\ 
2015-Jul-16 & R$_\mathrm{C}$& 81.041 & 44.69$\pm$1.07 & 158.45$\pm$0.65 & 44.67$\pm$1.07 & -0.93$\pm$0.65 \\ 
2015-Jul-16 & R$_\mathrm{C}$ & 81.042 & 44.49$\pm$0.90 & 157.23$\pm$0.59 & 44.37$\pm$0.90 & -2.15$\pm$0.59 \\ 
2015-Jul-16 & R$_\mathrm{C}$ & 81.044 & 44.48$\pm$0.90 & 158.03$\pm$0.56 & 44.43$\pm$0.90 & -1.36$\pm$0.56 \\ 
 \multicolumn{3}{c} {Weighted Mean} & 44.58$\pm$0.49 & 157.73$\pm$0.31 & 44.49$\pm$0.49 & -1.66$\pm$0.31 \\ 
 \hline
2015-Jul-16 & V & 81.046 & 44.72$\pm$1.09 & 157.58$\pm$0.70 & 44.63$\pm$1.09 & -1.81$\pm$0.70 \\ 
2015-Jul-16 & V & 81.047 & 44.86$\pm$1.06 & 156.75$\pm$0.71 & 44.66$\pm$1.06 & -2.65$\pm$0.71 \\ 
2015-Jul-16 & V & 81.049 & 42.06$\pm$1.14 & 158.88$\pm$0.72 & 42.05$\pm$1.14 & -0.52$\pm$0.72 \\ 
 \multicolumn{3}{c} {Weighted Mean}  & 43.95$\pm$0.63 & 157.72$\pm$0.41 & 43.85$\pm$0.63 & -1.68$\pm$0.41 \\ 
 \hline
2015-Jul-16 & R$_\mathrm{C}$ & 81.050 & 44.44$\pm$1.01 & 158.31$\pm$0.62 & 44.41$\pm$1.01 & -1.08$\pm$0.62 \\ 
2015-Jul-16 & R$_\mathrm{C}$ & 81.052 & 42.62$\pm$1.24 & 159.33$\pm$0.76 & 42.62$\pm$1.24 & -0.07$\pm$0.76 \\ 
2015-Jul-16 & R$_\mathrm{C}$ & 81.053 & 44.71$\pm$1.60 & 159.39$\pm$0.92 & 44.71$\pm$1.60 & -0.01$\pm$0.92 \\ 
 \multicolumn{3}{c} {Weighted Mean}  & 43.90$\pm$0.70 & 158.86$\pm$0.43 & 43.89$\pm$0.70 & -0.53$\pm$0.43 \\  
\hline
\hline
\end{tabular}
\tablefoot{
\tablefoottext{a}{Employed filters (V: center 545, width 87 nm R$_\mathrm{C}$: center 641, width 149 nm).}
\tablefoottext{b}{Solar phase angle.}
\tablefoottext{c}{Degree of linear polarization.}
\tablefoottext{d}{Position angle of polarization (in degrees E of N).}
\tablefoottext{e}{Degree of linear polarization with respect to the scattering plane.}
\tablefoottext{f}{Position angle of the polarization with respect to the scattering plane (in degrees E of N).}
}
\end{table*}
\section{Results}
\subsection{Linear polarization degree}
We found that 1998 KU$_\mathrm{2}$ with its very low geometric albedo ($p_\mathrm{V}$) exhibits enormously high polarization degrees ($P_\mathrm{r}$), that is, $P_\mathrm{r}$ = 44.6\% $\pm$ 0.5\% and 43.9\% $\pm$ 0.7\% at  UT 2015 July 16 for the R$_\mathrm{C}$
band and  44.0\% $\pm$ 0.6\% at UT 2015 July 16 for the V band at the phase angle ($\alpha$) of 81.0\degr (see Table~\ref{table2}). This finding is unexpected and remarkable although an object
with such a very low albedo has not been observed at larger phase angles in the past. These values are equal to approximately 1.5 times the maximum values of well-known high-polarization comets (typically, $P_\mathrm{r}\sim$ 28\% for $p_\mathrm{V}\sim$ 0.05) and more than four times as high as the $P_\mathrm{max}$ of S-complex asteroids. Although new research has revealed that Phaethon has a higher polarization degree ($P_\mathrm{r}$ = 50.1\%) at 106.5\degr \citep{Ito:2017}, the trend line of 1998 KU$_\mathrm{2}$ displays a polarization degree that is even higher than that of Phaethon at the studied phase angles, thus suggesting that the asteroid exhibits the highest polarization degree of the known airless objects in the solar system. We also find that the polarization degrees are less dependent on the wavelengths. Certain S-complex NEAs, including (1566) Icarus, (4179) Toutatis, and (23187) 2000 PN$_\mathrm{9}$, show red polarimetric colors \citep{Ishiguro:1997,Ishiguro:2017, Belskaya:2009b}. The neutral polarimetric color of 1998 KU$_\mathrm{2}$ can be explained by the inherent aspects of either low-albedo asteroids or 1998 KU$_\mathrm{2}$.

\begin{figure}
\centering
\includegraphics[width=\hsize]{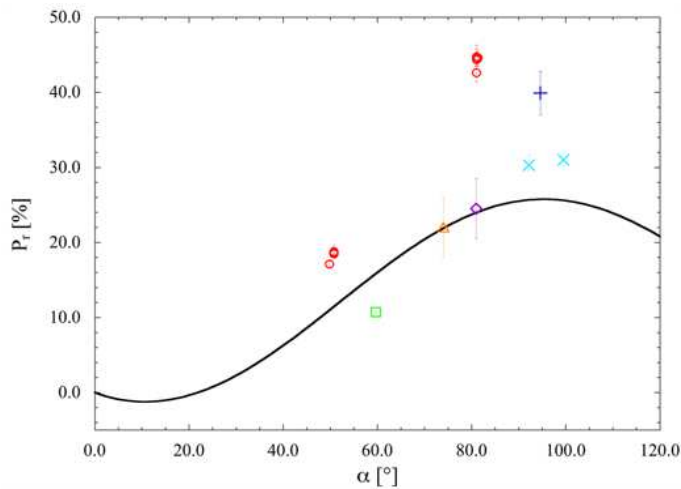}
\caption{Comparison of linear polarization degrees of 1998 KU$_\mathrm{2}$ and other bodies in the solar system in the red region (about 650 $\pm$ 50 nm). 
The open circles indicate data corresponding to 1998 KU$_\mathrm{2}$. Other symbols denote data for Ra-Shalom (open square), Phobos (open diamond), Deimos (open triangle), 2P/Encke (plus), and 209P/LINEAR (crosses). The solid line corresponds to high-polarization comets \citep{Levasseur-Regourd:1996}.}
\label{fig1}
 \end{figure}

Uncommonly high polarization degrees are also present at other phase angles: $P_\mathrm{r}$ = 17.1\% $\pm$ 0.4\% at $\alpha$ = 49.8\degr \ and  $P_\mathrm{r}$ = 18.6\% $\pm$ 0.3\% at $\alpha$ = 50.7\degr.  Figure~\ref{fig1} compares the polarization degrees of 1998 KU$_\mathrm{2}$ and those of a few airless objects with similar taxonomic type: (2100) Ra-Shalom  \citep[$p_\mathrm{V}$ = 0.08--0.14,][]{Harris:1998} and the two Martian satellites,  \citep[$p_\mathrm{V}$ = 0.07,][]{Zellner:1974b} and Deimos \citep[$p_\mathrm{V}$ = 0.07,][]{Thomas:1996}. The polarization degree of Ra-Shalom, which was reported as $P$= 10.7\% at $\alpha$ = 60\degr \citep{Kiselev:1999},  is obviously lower than that of 1998 KU$_\mathrm{2}$ at around $\alpha$ = 50\degr, although Ra-Shalom is classified as a C-complex asteroid, similar to 1998 KU$_\mathrm{2}$. The low-albedo satellites of Mars, Phobos and Deimos, show polarization values of $P$ = 24.5\% $\pm$ 4\% at $\alpha$ = 81\degr  and $P$ = 22\% $\pm$ 4\% at $\alpha$ = 74\degr \ in the orange domain (570 nm;  \citep{Noland:1973}. The polarization degree difference between 1998 KU$_\mathrm{2}$ and the Martian satellites is obvious because the polarimetric color trend between the R$_\mathrm{C}$  and V bands exhibits little difference within their errors. Cometary nuclei (typical $p_\mathrm{V}\sim$ 0.05) are identified with dark asteroids in terms of the albedo; their polarization degrees at large phase angles have been reported as $P_\mathrm{r}$ = 39.9\% $\pm$ 2.9\% at $\alpha$ = 94.6\degr for 2P/Encke \citep{Jockers:2005} and $P_\mathrm{r}$ = 31.0\% $^{+1.0}_{-0.7}$\% at $\alpha$ = 99.5\degr  for 209P/LINEAR \citep{Kuroda:2015}. None of these values exceeds the degree of polarization of 1998 KU$_\mathrm{2}$. To summarize, based on the geometric albedo, it is more reasonable to assume that the polarization degree is largely dependent on the surface albedo, as first advocated by \citet{Umow:1905}.
\subsection{Estimation and implication of polarimetric parameters}
Because this is the first time that polarimetric measurements of a dark asteroid have been performed at large phase angles, there is no precedent that can be directly compared with our results. Certain major polarimetric parameters (i.e., the polarization slope and inverse angle), the majority of which were acquired at phase angles smaller than 30\degr, have been known to exhibit typical values of the corresponding taxonomic types according to previous statistical studies \citep{Belskaya:2005,Belskaya:2017, Gil-Hutton:2011,Gil-Hutton:2012, Canada-Assandri:2012, Cellino:2012}. 

\begin{figure}
\centering
\includegraphics[width=\hsize]{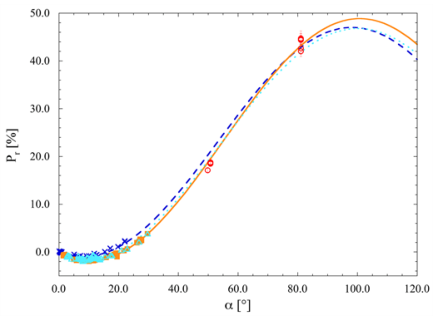}
\caption{
Polarization phase curve of 1998 KU$_\mathrm{2}$  along with those of three C-subgroup asteroids. Triple fit lines with the Lumme and Muinonen function (Eq.~(\ref{eq:LM})) indicate F-types (dashed line and crosses), Ch- and Cgh-types (solid line and filled squares), and other C-types (dotted line and open triangles) with geometric albedo < 0.1 \citep[][the unpublished data of Kiselev and Lupishko including the Asteroid Polarimetric Database \citep{Lupishko:2014}]{Zellner:1974a, Zellner:1976, Belskaya:1987, Belskaya:2009a, Hadamcik:2011, Gil-Hutton:2012, Nakayama:2000}.}
\label{fig2}
\end{figure}

Since 1998 KU$_\mathrm{2}$ is classified as an F-type \citep{Whiteley:2001} or Cb-type \citep{Binzel:2004} asteroid, we derived these polarimetric parameters to compensate for the missing data from the Asteroid Polarimetric Database  \citep{Lupishko:2014} and \citet{Hadamcik:2011}. According to \citet{Belskaya:2017}, the polarimetric slope parameter ($h$) and the inverse angle ($\alpha_\mathrm{0}$), which are defined by the ascending slope to the positive branch and the sign transition point, respectively, are useful parameters for distinguishing F- and Ch- and Cgh-type asteroids from other C-type asteroids. We selected the data of the polarization degrees in the red region ($\sim$R band) and low albedo ($p_\mathrm{V}$ < 0.1) for each taxonomic type, and we determined their polarimetric parameters by applying the Lumme and Muinonen function \citep{Goidet-Devel:1995, Penttila:2005} as
\begin{equation}
P(\alpha)=b\left(\sin \alpha\right)^{c_1} \left( \cos\left(0.5\alpha\right)\right)^{c_2} \sin\left(\alpha-\alpha_0\right),
 \label{eq:LM}
\end{equation}
\noindent
where $b$, $c_\mathrm{1}$, $c_\mathrm{2}$ , and $\alpha_\mathrm{0}$ denote positive constant parameters. By definition, the derivative of $P(\alpha)$ at $\alpha_\mathrm{0}$ represents the polarimetric slope (i.e., \begin{math} h = \left. \frac{\mathrm{d}P}{\mathrm{d}\alpha} \right|_{\alpha=\alpha_\mathrm{0}}\end{math}). Using the nonlinear least-squares fitting of this function, we obtained the polarimetric curves for each taxonomic type, as shown in Figure~\ref{fig2}. The curve based on the data set of Ch- and Cgh-types is smoothly coincident with the linear polarization curve of 1998 KU$_\mathrm{2}$; however, there is a slight mismatch between the curves from $\alpha$ = 49.8\degr~to $\alpha$ = 81.0\degr. As a result, we obtained the following polarimetric parameters: h = 0.330\%/\degr $\pm$ 0.003\%/\degr, $\alpha_\mathrm{0}$ = 21.2\degr $\pm$ 0.2\degr,  and $P_\mathrm{max}$ = 48.8\% $\pm$ 5.2\% for the Ch- and Cgh-type asteroids. These estimates, along with those of the others, are summarized in Table~\ref{table3}. All three types (F-, Ch- and Cgh-, and other C-types) present similar inverse angles as the corresponding mean values in previous studies \citep{Gil-Hutton:2012, Belskaya:2017}. In contrast, there are obvious differences in the polarimetric slope between our result and those of previous studies, and this difference is particularly large for the F-type (see the dashed line in Fig.~\ref{fig2}). From the polarimetric point of view, the taxonomic type of 1998 KU$_\mathrm{2}$ should not be regarded as F-type (also see section 4.1). The maximum polarization degree is about 47--49\%, with a large uncertainty arising from the extrapolation region for the limited data. Therefore, our result at $\alpha$ = 81.0\degr ~should pinpoint at least the lower limit of $P_\mathrm{max}$.

The traditional relation between the geometric albedo ($p_\mathrm{V}$) and the polarimetric slope ($h$), which is based on studies of scattering properties  \citep{Zellner:1974a, Dollfus:1989}, is known as the slope-albedo law, and it can be expressed as the following equation:
\begin{equation}
         log\ \mbox{$p_\mathrm{V}(h)$}=C_1\ log\ h + C_2~,
 \label{eq:Pv-h}
\end{equation}
\noindent
where $C_\mathrm{1}$ and $C_\mathrm{2}$ represent constants. Upon setting $C_\mathrm{1}$  = --1.111 $\pm$ 0.031 and  $C_\mathrm{2}$ = --1.781$\pm$ 0.025 \citep{Cellino:2015}, or $C_\mathrm{1}$ = --1.207 $\pm$ 0.067 and $C_\mathrm{2}$ = --1.892 $\pm$ 0.141 \citep{Masiero:2012}, the geometric albedos of 1998 KU$_\mathrm{2}$ are calculated as $p_\mathrm{V}(h)$ = 0.057 $\pm$ 0.004 or $p_\mathrm{V}(h)$ = 0.049 $\pm$ 0.011 for the above obtained slope values. 
On the other hand, the polarimetric slopes are computed as 0.93 \%/\degr \ or 0.76 \%/\degr \ when the geometric albedo of 0.018 derived from thermal-infrared observations \citep{Mainzer:2011} is substituted in Equation (\ref{eq:Pv-h}). This mismatch may be attributed to systematic uncertainties associated with the rotational changes in optical and thermal fluxes, since it is considered that 1998 KU$_\mathrm{2}$ has a large amplitude (1.35 $\pm$ 0.2 mag.) and a long rotational period (125 $\pm$ 5 hours) \citep{Warner:2016}. However, this is probably insignificant because the thermal-infrared data of NEOWISE \citep[e.g.,][]{Nugent:2016} were provided with $\sim$80 \% of the rotational phase coverage. Therefore, the very low albedo of 1998 KU$_\mathrm{2}$ is inconsistent with the albedos determined from the above polarimetric empirical relation (Eq.~(\ref{eq:Pv-h})), while the nonlinear trend between $p_\mathrm{V}(h)$ and $h$ appears to agree rather well with the laboratory measurements of pulverized (50--340 $\mu$m) terrestrial rocks \citep{Geake:1986}. 
\citet{Cellino:2015} described the existence of extremely low-albedo ($p_\mathrm{V}$ < 0.02) asteroids as derived by thermal observations as an open question, because such an albedo places some stringent constraints on the mineral composition of the
surface. It is also noteworthy that Barbarian asteroids with the unusual polarimetric behavior have a tendency to exhibit long spin rates \citep{Masiero:2009, Cellino:2014, Devogele:2017}. Our polarimetric finding demonstrates the possible presence of such extremely low-albedo asteroids, and we therefore need to reconsider the polarimetric behavior of these asteroids for the basic insights that they may offer. 
\begin{table*}
\caption{\label{table3}Estimates of polarimetric parameters}
\centering
\begin{tabular}{lcccccc}
\hline\hline
Taxonomic type &  \multicolumn{2}{c}  {$h\pm\sigma_h$\tablefootmark{a}} & \multicolumn{2}{c}  {$\alpha_0\pm\sigma_{\alpha_0}$\tablefootmark{b}} & $P_\mathrm{max}\pm\sigma_{P_\mathrm{max}}\tablefootmark{c}$ & $\chi_\nu^\mathrm{2}$\,\tablefootmark{d} \\
&  \multicolumn{2}{c} {[\%/\degr]} &  \multicolumn{2}{c} {[\degr]} &  [\%] &  \\
 & this work & previous work\tablefootmark{e}  & this work & previous work\tablefootmark{e}   &  &  \\ 
\hline
F & 0.234$\pm$0.005 & 0.608$\pm$0.193 & 16.2$\pm$0.3 & 15.7$\pm$0.2 & 47.0$\pm$7.6 & 11.6 \\ 
Ch \& Cgh & 0.330$\pm$0.003 & 0.440$\pm$0.050 & 21.2$\pm$0.2 & 21.3$\pm$0.1 & 48.8$\pm$5.2 & 4.2 \\ 
Other C & 0.307$\pm$0.005 & 0.387$\pm$0.037 & 19.5$\pm$0.4 & 19.4$\pm$0.1 & 47.0$\pm$7.6 & 16.6 \\ 
\hline
\hline
\end{tabular}
\tablefoot{
\tablefoottext{a}{Polarimetric slope parameter.}
\tablefoottext{b}{Inverse angle.}
\tablefoottext{c}{Maximum of polarization degree.}
\tablefoottext{d}{Reduced chi-square value.}
\tablefoottext{e}{\citet{Belskaya:2017}}
}
\end{table*}
\subsection{Cometary activity}
Cometary activity (coma and/or tails) for 1998 KU$_\mathrm{2}$ was undetectable in our observation runs, even though a percentage of dark asteroids with primitive composition are regarded as objects of cometary origin \citep{Kim:2014}. The abundance of such objects is 8\% $\pm$ 5\% \citep{DeMeo:2008} and 4\% \citep{Fernandez:2005} in NEAs under certain dynamical criteria (i.e., Tisserand's parameter:  T$_\mathrm{j}$). The polarization degrees during the disruption of the cometary nucleus with jet-like features were known to be higher than the degree of the whole coma, and these retained positive values through all the phase angles \citep{Hadamcik:2003}. 1998 KU$_\mathrm{2}$ has an orbit that is more typical of asteroids than comets (based on the Tisserand's parameter with respect to Jupiter T$_\mathrm{j}$ = 3.40, where T$_\mathrm{j}$ < 3 indicates a comet-like object, while T$_\mathrm{j}$ > 3 indicates an asteroid-like
object). However, since some asteroids have displayed cometary activity \citep[see e.g.,][]{Jewitt:2012}, we examined the dust environment as described below.

\begin{figure}
\centering
\includegraphics[width=\hsize]{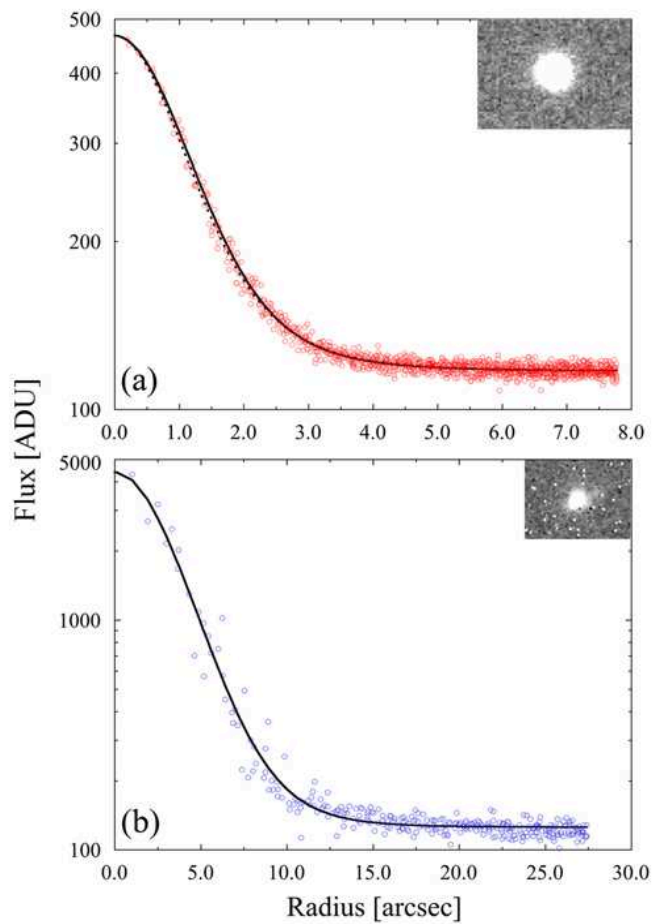}
\caption{Radial profiles obtained from observations made in (a) this work at UT 2015 June 12 in the R$_\mathrm{C}$ band and (b) NEOWISE at UT 2015 July 08 in the W2(4.6 $\mu$m) band. The open circles indicate the extracted data points of the radial profile of 1998 KU$_\mathrm{2}$. The solid and dotted lines correspond to Moffat's PSF of the star scaled to the flux of 1998 KU$_\mathrm{2}$.}
\label{fig3}
 \end{figure}

We first describe a simple approach based on comparisons with the point-spread function (PSF) profiles of 1998 KU$_\mathrm{2}$ and nearby background stars. Twelve frames (three cycles in our polarimetric observations) were aligned to each pixel position of 1998 KU$_\mathrm{2}$ and then combined to generate a single frame. The radial profile of 1998 KU$_\mathrm{2}$ was measured on its coadded frame using the IRAF pradprof task. Owing to the non-sidereal tracking of the telescope, the profiles of the trailing reference stars were extracted only as perpendicular components of the trail direction. Figure~\ref{fig3}a presents the scaled radial profiles of 1998 KU$_\mathrm{2}$ and two reference stars. Since all PSFs exhibit good consistency, we can conclude that no source extension appeared during our observations.

The similarly low-albedo ($p_\mathrm{V}$ = 0.03) NEA (3552) Don Quixote, which was speculated to be an extinct or dormant comet on the basis of its orbit and spectroscopic features, presented a coma and a tail only in the 4.5-$\mu$m band of Spitzer/IRAC \citep{Mommert:2014}. We attempted to detect the cometary activity of 1998 KU$_\mathrm{2}$  using the image of the 4.6 $\mu$m band of NEOWISE \citep{Mainzer:2014}, but its radial profile conformed to the scaled stellar PSF, as in the case of our observation (Fig.~\ref{fig3}b).
Moreover, photometric observations of 1998 KU$_\mathrm{2}$ during 2015 were performed by \citet{Clark:2016} and \citet{Warner:2016}, and neither author determined its cometary signature. From this evidence, we conclude that the surface of 1998 KU$_\mathrm{2}$ is essentially similar to those of typical dark asteroids. 
\section{Discussions}
To interpret the significantly high degree of linear polarization of 1998 KU$_\mathrm{2}$, in this section, we propose a feasible surface whose condition is set to satisfy our collateral facts (e.g., polarimetric curve similar to Ch- and Cgh-type asteroids, non-active asteroid), spectral features, and thermal-infrared results.

\subsection{Spectral features}
\begin{figure}
\centering
\includegraphics[width=\hsize]{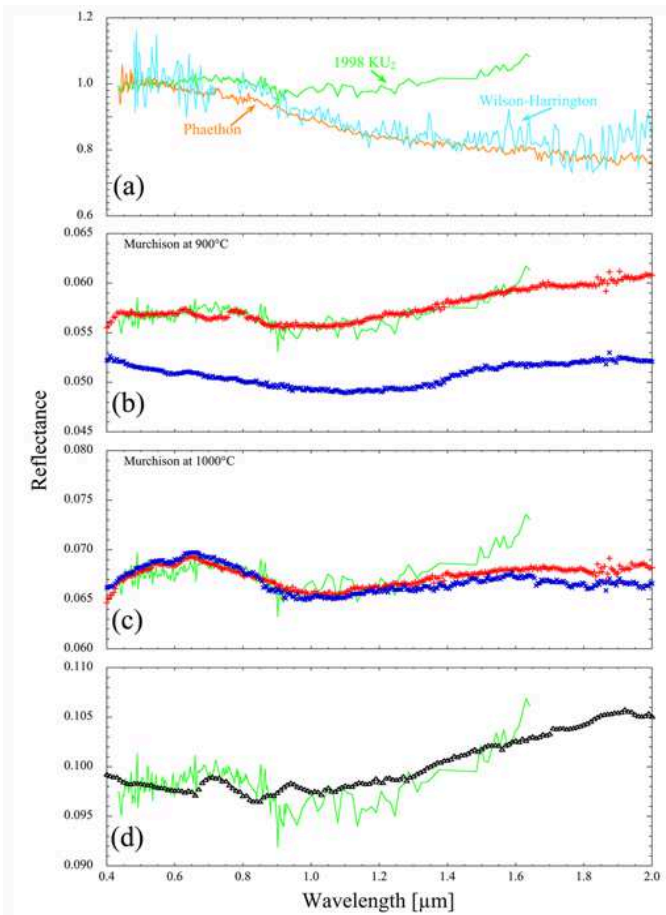}
\caption{
Comparison of spectral features of asteroids and meteorites in the visible and near-infrared regions. 1998 KU$_\mathrm{2}$ (solid lines) is compared with (a) other F-type near-Earth asteroids, Phaethon and Wilson-Harrington, (b)--(c) the heated Murchison meteorites at 900 $^\circ$C or 1000 $^\circ$C  (the pluses and crosses indicate a grain size of < 0.63 $\mu$m and a grain size range of 63--125 $\mu$m, respectively), (d) the Almahata Sitta meteorite (triangle). Asteroid and meteorite spectra were taken from SMASS\protect \footnotemark[3], \citet{Ishiguro:2011} and RELAB\protect \footnotemark[4], M4AST\protect \footnotemark[5] \citep{Popescu:2012} was used as the matching tool.}
\label{fig4}
\end{figure}

The taxonomic type of 1998 KU$_\mathrm{2}$, Cb-type, was determined by examining the visible spectrum \citep[e.g., Bus taxonomy,][]{Bus:2002}. The spectrum up to 1.6 $\mu$m was obtained through the SMASS survey \citep{Binzel:2004}; however, the data have not been used so far for the further study of taxonomic classification. Since the visible spectra of C-complex asteroids are featureless, we should consider 1998 KU$_\mathrm{2}$ as a C-type asteroid rather than as belonging to other subgroups upon applying the
Bus-DeMeo taxonomy \citep{DeMeo:2009}. Our polarimetric result, which implies that 1998 KU$_\mathrm{2}$ is not an F-type asteroid (see section 3.1 and the prediction with the dashed line in Fig.~\ref{fig2}) is consistent with this classification even though the criteria used in the Bus and Bus-DeMeo classification cannot distinguish F-type asteroids. Meanwhile, two other F-type NEAs, (3200) Phaethon \citep{Hicks:1998} and (4015) Wilson-Harrington \citep{Tholen:1989}, which exhibit the spectral trend of reflectance decrease with increasing wavelength, are obviously different from that of 1998 KU$_\mathrm{2}$ (see Fig.~\ref{fig4}a). 

When comparing our results with meteorite spectra from the RELAB database \citep{Pieters:2004}, we determined that the continuous spectral trend of 1998 KU$_\mathrm{2}$ corresponds with that of the Murchison (CM) meteorite heated at 900$^\circ$C, except for an absorption feature at around 0.7 $\mu$m (Fig.~\ref{fig4}b), while the absorption position and shape are fairly similar to the corresponding ones of the heated sample of the Murchison meteorite at 1000$^\circ$C (Fig.~\ref{fig4}c). These spectra were measured from meteorite grains smaller than 63 $\mu$m, and partially different patterns were obtained with grain sizes of 63--125 $\mu$m (Fig.~\ref{fig4}b and \ref{fig4}c). Although a very low albedo cannot be produced in these laboratory samples, we speculate that 1998 KU$_\mathrm{2}$ comprises a regolith with a grain size of several tens of microns exhibiting Murchison-like mineral composition corresponding to heating and/or thermal metamorphism between 900$^\circ$C and 1000$^\circ$C. In general, such thermal alterations may be interpreted with the well-known space-weathering product, which is attributed to micrometeorite bombardment or solar wind sputtering, because such weathering yields a darker surface and the absorption features appear weaker, while the spectral curvature appears "redder" in ordinary chondrites \citep{Sasaki:2001, Brunetto:2005}. However, some experimental results regarding carbonaceous chondrites have indicated no regular pattern basis  at this point \citep[][and references therein]{Kaluna:2017}.

Meanwhile, a slightly bluish flat spectrum characterized by features of small absorption and low albedo corresponds to the typical F-type asteroid \citep{Tholen:1984}. The taxonomic features and orbit of 1998 KU$_\mathrm{2}$ are similar to those of 2008 TC$_\mathrm{3}$ , which collided with Earth, and the former was known as one of the candidates for the parent body \citep{Jenniskens:2009, Jenniskens:2010}. The fallen meteorite, which was named the Almahata Sitta meteorite, consisted of various mineralogical types associated with different meteorite groups (Bischoff et al., 2010; Kohout et al., 2010). The majority component is classified as anomalous polymict ureilite in primitive achondrites \citep{Bischoff:2010, Kohout:2010}. A similar spectrum has also been found in the RELAB database (Almahata Sitta 4 chip lighter face, as shown in Figure~\ref{fig4}d), and its albedo tends to be bright \citep{Hiroi:2010}.  The albedo and spectra obtained in the laboratory measurements, show an obscurity in the parental relationship between 1998 KU$_\mathrm{2}$ and 2008 TC$_\mathrm{3}$. As a clue, the unique polarimetric character of F-type asteroid may provide useful information in identifying the parent body of carbonaceous chondrites and/or primitive achondrites.
\footnotetext[3]{http://smass.mit.edu/}
\footnotetext[4]{http://www.planetary.brown.edu/relab/}
\footnotetext[5]{http://m4ast.imcce.fr/}
\subsection{Estimation of surface regolith structure}
\begin{figure}
\centering
\includegraphics[width=\hsize]{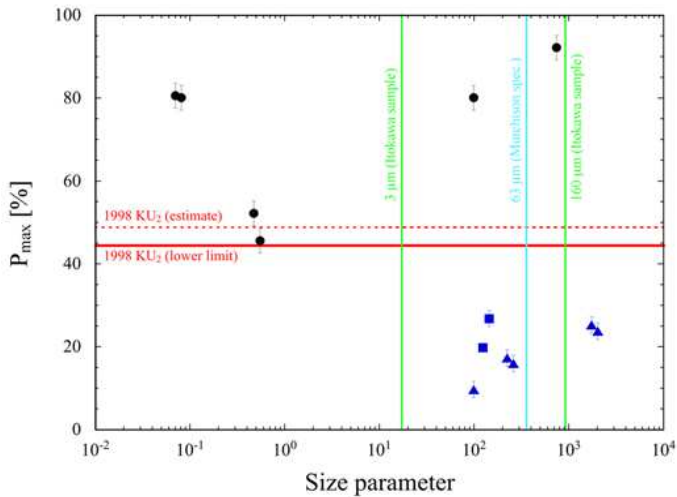}
\caption{Dependence of the particle size on parameter of the polarization maximum ($P_\mathrm{max}$). The filled circles indicate size-sorted carbon samples \citep{Hadamcik:2009}. The filled squares and filled triangles denote two size-sorted carbonaceous chondrites \citep[e.g., Orgueil and Allende meteorites,][]{Worms:2000, Hadamcik:2011}, respectively. The 3 $\mu$m and 160 $\mu$m lines are qualified from the Itokawa sample limits. The 63 $\mu$m line originates from the spectral resemblance of the heated Murchison meteorite (section 4.1). The $P_\mathrm{max}$ values of 1998 KU$_\mathrm{2}$ correspond to the lower limit from our observation (solid line) and the estimate made with the Lumme and Muinonen function (Eq.~(\ref{eq:LM})) in Fig.~\ref{fig2} (dashed line).}
\label{fig5}
\end{figure}

1998 KU$_\mathrm{2}$ exhibited an unprecedentedly high degree of polarization, which implies the existence of an asteroid with very low albedo (i.e., 0.018, see section 3.2). In section 4.1, we stated that the surface composition of this asteroid can be realized with thermal alteration of known carbonaceous chondrite based on the spectroscopic analysis. Therefore, we suspect that some specific surface regolith structures (such as size or porosity) can be attributed to the polarimetric peculiarity and geometric albedo. To estimate the effective regolith particle size, we applied an empirical relation between $P_\mathrm{max}$ and the particle size derived from previous studies for laboratory samples \citep{Worms:2000, Hadamcik:2009, Hadamcik:2011}. Figure~\ref{fig5} presents the $P_\mathrm{max}$ targeted at two carbonaceous chondrites (Orgueil and Allende meteorites) and the amorphous carbon for each particle size. These polarization degrees were measured under the sample-deposited condition, and the size parameter was defined as X = $\pi\mathrm{d}/\lambda$~\citep{Bohren:1983}, where d and $\lambda$ represent the particle equivalent diameter and wavelength, respectively.

When d > $\lambda$, the relationship that indicates that a larger particle size affords higher polarization degrees does not appear as a unique trend considering the conditions and features of the particle components. The $P_\mathrm{max}$ values corresponding to d < $\lambda$, which were obtained from 1--10 $\mu$m sized fluffy aggregates composed of tiny carbon grains, appear to increase with smaller grain sizes. The $P_\mathrm{max}$ of Orgueil and Allende do not correspond to the estimated $P_\mathrm{max}$ of 1998 KU$_\mathrm{2}$ over any size range; in comparison with $P_\mathrm{max}$ for a given size, that of Orgueil is higher. Amorphous carbon with high $P_\mathrm{max}$ is one of the representative opaque materials among carbonaceous chondrites. Thus, the degree of polarization and albedo may originate in the carbon content, because bulk carbon abundances including other carbon-bearing species were reported as 4.88 wt\% for Orgueil and 0.27 wt\% for Allende \citep{Pearson:2006}. 

As mentioned in the previous section, the spectral results of 1998 KU$_\mathrm{2}$ imply a particle size smaller than 63 $\mu$m. Since dust particles obtained from NEA (25143) Itokawa have a nominal size range from a few microns to 160 $\mu$m \citep{Nakamura:2011, Tsuchiyama:2011}, similar-sized regolith particles are also assumed to exist on asteroids with sizes of several kilometers. If the $P_\mathrm{max}$ of Orgueil (19.8\% at X = 99--149 and 26.8\% at X = 116--173) adjusts to nearly that of 1998 KU$_\mathrm{2}$ with the addition of only 20 $\mu$m sized compact carbon (e.g., $P_\mathrm{max}$ 80.1\% at X = 50--150), the mixing ratio of carbon is required to be at least about 41\% in this size range. Because the bulk abundance of carbon becomes significantly higher than the original abundance, it is probably unreasonable to assume that compact carbon particles with sizes > 20 $\mu$m exist on 1998 KU$_\mathrm{2}$. 

We instead suggest the presence of micron-sized fluffy aggregates with a diameter of several tens of nanometers that comprise carbon grains, because this yields a similarly high $P_\mathrm{max}$ (e.g., $P_\mathrm{max}\sim$ 80.1\% for the grain size parameter range of 0.052--0.110) as that of 20 $\mu$m sized compact carbon particles, and the carbon requirement is significantly lower than for the case of compact particles. Furthermore, the extremely low albedo  (0.001 $\pm$ 0.001) of this porous aggregate is particularly noteworthy; a decrease in the geometric albedo can be expected, even if the albedo of the compact particles was not described by \citet{Hadamcik:2009}. According to the results of an experiment resembling the impact reaction on the asteroid surface, many types of carbon nanoclusters can be produced in the gas environment \citep{Mieno:2011}. Thus, we expect that tiny carbon grains may be generated when interplanetary dust and meteorites collide with organic material on the surface layer. Such a bombardment heating process is consistent with the spectral results discussed in section 4.1. 

Another aspect describing the regolith structure involves thermal inertia, which has been regarded as a sensitive indicator of the typical regolith particle size \citep{Gundlach:2013}.  The thermal inertia of 1998 KU$_\mathrm{2}$, which is not determined from previous thermal observations, can be estimated to be hundreds to thousands of Jm$^\mathrm{-2}$s$^\mathrm{-1/2}$K$^\mathrm{-1}$. The lower limit is a typical value for a beaming parameter of 0.901 at $\alpha$ = 16.8\degr~and a diameter of 4.7 km \citep{Mainzer:2011} on the analogy of a mean thermal inertia of 200 $\pm$ 40 Jm$^\mathrm{-2}$s$^\mathrm{-1/2}$K$^\mathrm{-1}$\citep{Delbo:2007} for kilometer-sized NEAs, while the upper limit is an extrapolation value associated with slow-spin NEAs \citep{Harris:2016}. The regolith grains derived from this range are inferred to have an effective size of gravel (several millimeters to tens of centimeters) on this surface. Although there are significant differences with our polarimetric constraint, this can be explained on the basis of skin depth. In this case, the thermal skin depth whose equation constitutes the mass density, specific heat capacity, and thermal conductivity \citep{Spencer:1989} is calculated deeper than a few centimeters. Thus, the optical results reflect the presence of a shallow surficial deposit, whereas the thermal result represents the subsoil features of the top few centimeters. 

Considering our polarimetric results and other studies of 1998 KU$_\mathrm{2}$, as regards a probable regolith structure, we finally conclude that micron-sized fluffy aggregates composed of mainly nano-sized carbon grains are deposited on the gravel-sized material layer. These fluffy aggregates correspond to the shape of interplanetary dust particles (IDPs) acquired in the stratosphere \citep{Brownlee:1985}. Some IDPs have an albedo of around 0.02 \citep{Bradley:1996}. The surface layers of some C-complex asteroids have spectral properties compatible with those of IDPs \citep{Vernazza:2015, Vernazza:2017, Hasegawa:2017}. These facts support the existence of micron-sized fluffy aggregates with nano-sized carbon grains on the surface of 1998 KU$_\mathrm{2}$.
\section{Summary}
We conducted multiband polarimetric observations of a very low-albedo NEA, (152679) 1998 KU$_\mathrm{2}$, at phase angles of 49.8\degr, 50.7\degr, and 81.0\degr. We report the following findings:
\begin{enumerate}
 \item Significantly high linear polarizations were detected at each phase angle when compared with those observed in past studies.
 \item There is almost no difference in the polarization degrees for the V and R$_\mathrm{C}$ bands at $\alpha$ = 81.0\degr.
 \item 1998 KU$_\mathrm{2}$ presents polarimetric characteristics most similar to Ch- and Cgh-type asteroids. Its $P_\mathrm{max}$ as obtained from extrapolation with the empirical equation was estimated as 48.8\%$\pm$5.2\%.
 \item No cometary activity was detected in the optical and mid-infrared regions.
 \item Interpreting spectroscopic and polarimetric measurements based on previous laboratory studies, we estimate that 1998 KU2 most likely has a regolith structure consisting of micron-sized fluffy particles with nano-sized carbon grains deposited on top of the surface layer.
\end{enumerate}

\begin{acknowledgements}

The authors would like to thank Dr. Joe Masiero for enhancing the clarity of the manuscript. We also thank Mr. Yoonsoo P. Bach for providing helpful information about the thermal model.
Mr. Shuhei Goda helped us with the first installation of the polarimetric mode.  MI was supported by a National Research Foundation of Korea (NRF) grant funded by the Korean Government (MEST) (No. 2015R1D1A1A01060025). The work of SH was supported by the Hypervelocity Impact Facility (former facility name: the Space Plasma Laboratory), ISAS, JAXA. FU and SH were supported by JSPS KAKENHI Grant Numbers JP15K05277 and JP17K05636. This research was partially supported by the Optical \& Near-Infrared Astronomy Inter-University Cooperation Program, MEXT, of Japan.

This publication makes use of datasets from the Near-Earth Object Wide-field Infrared Survey Explorer (NEOWISE), which is a project of the Jet Propulsion Laboratory/California Institute of Technology. NEOWISE is funded by the National Aeronautics and Space Administration

Part of the data used in this publication was obtained and made available by The MIT-UH-IRTF Joint Campaign for NEO Reconnaissance. The IRTF is operated by the University of Hawaii under Cooperative Agreement no. NCC 5-538 with the National Aeronautics and Space Administration, Office of Space Science, Planetary Astronomy Program. The MIT component of this work is supported by NASA grant 09-NEOO009-0001, and by the National Science Foundation under Grants Nos. 0506716 and 0907766.

\end{acknowledgements}

%
%


\bibliographystyle{aa} 
\bibliography{Kuroda2017} 

\begin{thebibliography}{94}
\expandafter\ifx\csname natexlab\endcsname\relax\def\natexlab#1{#1}\fi

\bibitem[{{Belskaya} {et~al.}(2015){Belskaya}, {Cellino}, {Gil-Hutton},
  {Muinonen}, \& {Shkuratov}}]{Belskaya:2015}
{Belskaya}, I., {Cellino}, A., {Gil-Hutton}, R., {Muinonen}, K., \&
  {Shkuratov}, Y. 2015, {Asteroid Polarimetry} ({University of Arizona Press}),
  151--163

\bibitem[{Belskaya {et~al.}(2009)Belskaya, Fornasier, \&
  Krugly}]{Belskaya:2009b}
Belskaya, I.~N., Fornasier, S., \& Krugly, Y.~N. 2009, \icarus, 201, 167

\bibitem[{Belskaya {et~al.}(2017)Belskaya, Fornasier, Tozzi, Gil-Hutton,
  Cellino, Antonyuk, Krugly, Dovgopol, \& Faggi}]{Belskaya:2017}
Belskaya, I.~N., Fornasier, S., Tozzi, G.~P., {et~al.} 2017, \icarus, 284, 30

\bibitem[{{Belskaya} {et~al.}(2009){Belskaya}, {Levasseur-Regourd}, {Cellino},
  {Efimov}, {Shakhovskoy}, {Hadamcik}, \& {Bendjoya}}]{Belskaya:2009a}
{Belskaya}, I.~N., {Levasseur-Regourd}, A.-C., {Cellino}, A., {et~al.} 2009,
  \icarus, 199, 97

\bibitem[{{Belskaya} {et~al.}(1987){Belskaya}, {Lupishko}, \&
  {Shakhovskoi}}]{Belskaya:1987}
{Belskaya}, I.~N., {Lupishko}, D.~F., \& {Shakhovskoi}, N.~M. 1987, \sovast,
  13, 219

\bibitem[{Belskaya {et~al.}(2005)Belskaya, Shkuratov, Efimov, Shakhovskoy,
  Gil-Hutton, Cellino, Zubko, Ovcharenko, Bondarenko, Shevchenko, Fornasier, \&
  Barbieri}]{Belskaya:2005}
Belskaya, I.~N., Shkuratov, Y.~G., Efimov, Y.~S., {et~al.} 2005, \icarus, 178,
  213

\bibitem[{Binzel {et~al.}(2004)Binzel, Rivkin, Stuart, Harris, Bus, \&
  Burbine}]{Binzel:2004}
Binzel, R.~P., Rivkin, A.~S., Stuart, J.~S., {et~al.} 2004, \icarus, 170, 259

\bibitem[{{Bischoff} {et~al.}(2010){Bischoff}, {Horstmann}, {Pack},
  {Laubenstein}, \& {Haberer}}]{Bischoff:2010}
{Bischoff}, A., {Horstmann}, M., {Pack}, A., {Laubenstein}, M., \& {Haberer},
  S. 2010, Meteoritics and Planetary Science, 45, 1638

\bibitem[{{Bohren} \& {Huffman}(1983)}]{Bohren:1983}
{Bohren}, C.~F. \& {Huffman}, D.~R. 1983, {Absorption and scattering of light
  by small particles} ({University of Arizona Press})

\bibitem[{{Bradley} {et~al.}(1996){Bradley}, {Keller}, {Brownlee}, \&
  {Thomas}}]{Bradley:1996}
{Bradley}, J.~P., {Keller}, L.~P., {Brownlee}, D.~E., \& {Thomas}, K.~L. 1996,
  Meteoritics and Planetary Science, 31, 394

\bibitem[{{Brownlee}(1985)}]{Brownlee:1985}
{Brownlee}, D.~E. 1985, Annual Review of Earth and Planetary Sciences, 13, 147

\bibitem[{{Brunetto} \& {Strazzulla}(2005)}]{Brunetto:2005}
{Brunetto}, R. \& {Strazzulla}, G. 2005, \icarus, 179, 265

\bibitem[{{Bus} \& {Binzel}(2002)}]{Bus:2002}
{Bus}, S.~J. \& {Binzel}, R.~P. 2002, \icarus, 158, 146

\bibitem[{Ca{\~n}ada-Assandri {et~al.}(2012)Ca{\~n}ada-Assandri, Gil-Hutton, \&
  Benavidez}]{Canada-Assandri:2012}
Ca{\~n}ada-Assandri, M., Gil-Hutton, R., \& Benavidez, P. 2012, \aap, 542, A11

\bibitem[{{Cellino} {et~al.}(2014){Cellino}, {Bagnulo}, {Tanga},
  {Novakovi{\'c}}, \& {Delb{\`o}}}]{Cellino:2014}
{Cellino}, A., {Bagnulo}, S., {Tanga}, P., {Novakovi{\'c}}, B., \& {Delb{\`o}},
  M. 2014, \mnras, 439, L75

\bibitem[{{Cellino} {et~al.}(2015){Cellino}, {Gil-Hutton}, \&
  {Belskaya}}]{Cellino:2015}
{Cellino}, A., {Gil-Hutton}, R., \& {Belskaya}, I.~N. 2015, {Asteroids}
  ({Cambridge University Press}), 360--378

\bibitem[{Cellino {et~al.}(2012)Cellino, Gil-Hutton, Dell'Oro, Bendjoya,
  Ca{\~n}ada-Assandri, \& di~Martino}]{Cellino:2012}
Cellino, A., Gil-Hutton, R., Dell'Oro, A., {et~al.} 2012, \jqsrt, 113, 2552

\bibitem[{Chernova {et~al.}(1993)Chernova, Kiselev, \& Jockers}]{Chernova:1993}
Chernova, G.~P., Kiselev, N.~N., \& Jockers, K. 1993, \icarus, 103, 144

\bibitem[{{Clark}(2016)}]{Clark:2016}
{Clark}, M. 2016, Minor Planet Bulletin, 43, 2

\bibitem[{Delb{\`o} {et~al.}(2007)Delb{\`o}, Cellino, \& Tedesco}]{Delbo:2007}
Delb{\`o}, M., Cellino, A., \& Tedesco, E.~F. 2007, \icarus, 188, 266

\bibitem[{{DeMeo} \& {Binzel}(2008)}]{DeMeo:2008}
{DeMeo}, F. \& {Binzel}, R.~P. 2008, \icarus, 194, 436

\bibitem[{Demeo {et~al.}(2009)Demeo, Binzel, Slivan, \& Bus}]{DeMeo:2009}
Demeo, F.~E., Binzel, R.~P., Slivan, S.~M., \& Bus, S.~J. 2009, \icarus, 202,
  160

\bibitem[{{DeMeo} \& {Carry}(2013)}]{DeMeo:2013}
{DeMeo}, F.~E. \& {Carry}, B. 2013, \icarus, 226, 723

\bibitem[{{Devog{\`e}le} {et~al.}(2017){Devog{\`e}le}, {Tanga}, {Bendjoya},
  {Rivet}, {Surdej}, {Hanus}, {Abe}, {Antonini}, {Artola}, {Audejean},
  {Behrend}, {Berski}, {Bosch}, {Bronikowska}, {Carbognani}, {Char}, {Kim},
  {Choi}, {Colazo}, {Coloma}, {Coward}, {Durkee}, {Erece}, {Forne}, {Hickson},
  {Hirsch}, {Horbowicz}, {Kami{\'n}ski}, {Kankiewicz}, {Kaplan}, {Kwiatkowski},
  {Konstanciak}, {Kruszewki}, {Kudak}, {Manzini}, {Moon}, {Marciniak},
  {Murawiecka}, {Nadolny}, {Og{\l}oza}, {Ortiz}, {Oszkiewicz}, {Pallares},
  {Peixinho}, {Poncy}, {Reyes}, {de los Reyes}, {Santana-Ros}, {Sobkowiak},
  {Pastor}, {Pilcher}, {Qui{\~n}ones}, {Trela}, \& {Vernet}}]{Devogele:2017}
{Devog{\`e}le}, M., {Tanga}, P., {Bendjoya}, P., {et~al.} 2017, \aap, accepted
  [\eprint[arXiv]{1707.07503}]

\bibitem[{{Dollfus} {et~al.}(1989){Dollfus}, {Wolff}, {Geake}, {Dougherty}, \&
  {Lupishko}}]{Dollfus:1989}
{Dollfus}, A., {Wolff}, M., {Geake}, J.~E., {Dougherty}, L.~M., \& {Lupishko},
  D.~F. 1989, {Photopolarimetry of asteroids} ({University of Arizona Press}),
  594--616

\bibitem[{{Fern{\'a}ndez} {et~al.}(2005){Fern{\'a}ndez}, {Jewitt}, \&
  {Sheppard}}]{Fernandez:2005}
{Fern{\'a}ndez}, Y.~R., {Jewitt}, D.~C., \& {Sheppard}, S.~S. 2005, \aj, 130,
  308

\bibitem[{Fornasier {et~al.}(2015)Fornasier, Belskaya, \&
  Perna}]{Fornasier:2015}
Fornasier, S., Belskaya, I.~N., \& Perna, D. 2015, \icarus, 250, 280

\bibitem[{Geake \& Dollfus(1986)}]{Geake:1986}
Geake, J.~E. \& Dollfus, A. 1986, \mnras, 218, 75

\bibitem[{Gil-Hutton \& Ca{\~n}ada-Assandri(2011)}]{Gil-Hutton:2011}
Gil-Hutton, R. \& Ca{\~n}ada-Assandri, M. 2011, \aap, 529, A86

\bibitem[{Gil-Hutton \& Ca{\~n}ada-Assandri(2012)}]{Gil-Hutton:2012}
Gil-Hutton, R. \& Ca{\~n}ada-Assandri, M. 2012, \aap, 539, A115

\bibitem[{{Goidet-Devel} {et~al.}(1995){Goidet-Devel}, {Renard}, \&
  {Levasseur-Regourd}}]{Goidet-Devel:1995}
{Goidet-Devel}, B., {Renard}, J.~B., \& {Levasseur-Regourd}, A.~C. 1995,
  \planss, 43, 779

\bibitem[{Gundlach \& Blum(2013)}]{Gundlach:2013}
Gundlach, B. \& Blum, J. 2013, \icarus, 223, 479

\bibitem[{Hadamcik \& Levasseur-Regourd(2003)}]{Hadamcik:2003}
Hadamcik, E. \& Levasseur-Regourd, A.~C. 2003, \jqsrt, 79-80, 661

\bibitem[{Hadamcik {et~al.}(2011)Hadamcik, Levasseur-Regourd, Renard, Lasue, \&
  Sen}]{Hadamcik:2011}
Hadamcik, E., Levasseur-Regourd, A.~C., Renard, J.~B., Lasue, J., \& Sen, A.~K.
  2011, \jqsrt, 112, 1881

\bibitem[{Hadamcik {et~al.}(2009)Hadamcik, Renard, Levasseur-Regourd, Lasue,
  Alcouffe, \& Francis}]{Hadamcik:2009}
Hadamcik, E., Renard, J.~B., Levasseur-Regourd, A.~C., {et~al.} 2009, \jqsrt,
  110, 1755

\bibitem[{{Hanu{\v s}} {et~al.}(2016){Hanu{\v s}}, {Delbo'},
  {Vokrouhlick{\'y}}, {Pravec}, {Emery}, {Al{\'{\i}}-Lagoa}, {Bolin},
  {Devog{\`e}le}, {Dyvig}, {Gal{\'a}d}, {Jedicke}, {Korno{\v s}}, {Ku{\v
  s}nir{\'a}k}, {Licandro}, {Reddy}, {Rivet}, {Vil{\'a}gi}, \&
  {Warner}}]{Hanus:2016}
{Hanu{\v s}}, J., {Delbo'}, M., {Vokrouhlick{\'y}}, D., {et~al.} 2016, \aap,
  592, A34

\bibitem[{Harris(1998)}]{Harris:1998}
Harris, A.~W. 1998, \icarus, 131, 291

\bibitem[{{Harris} \& {Drube}(2016)}]{Harris:2016}
{Harris}, A.~W. \& {Drube}, L. 2016, \apj, 832, 127

\bibitem[{Hasegawa {et~al.}(2017)Hasegawa, Kuroda, Yanagisawa, \&
  Usui}]{Hasegawa:2017}
Hasegawa, S., Kuroda, D., Yanagisawa, K., \& Usui, F. 2017, \pasj, in press,
  psx117

\bibitem[{{Hicks} {et~al.}(1998){Hicks}, {Fink}, \& {Grundy}}]{Hicks:1998}
{Hicks}, M.~D., {Fink}, U., \& {Grundy}, W.~M. 1998, \icarus, 133, 69

\bibitem[{{Hiroi} {et~al.}(2010){Hiroi}, {Jenniskens}, {Bishop}, {Shatir},
  {Kudoda}, \& {Shaddad}}]{Hiroi:2010}
{Hiroi}, T., {Jenniskens}, P., {Bishop}, J.~L., {et~al.} 2010, Meteoritics and
  Planetary Science, 45, 1836

\bibitem[{{Ishiguro} {et~al.}(2011){Ishiguro}, {Ham}, {Tholen}, {Elliott},
  {Micheli}, {Niwa}, {Sakamoto}, {Matsuda}, {Urakawa}, {Yoshimoto}, {Sarugaku},
  {Usui}, {Hasegawa}, {Iwata}, {Ozaki}, {Kuroda}, \& {Ootsubo}}]{Ishiguro:2011}
{Ishiguro}, M., {Ham}, J.-B., {Tholen}, D.~J., {et~al.} 2011, \apj, 726, 101

\bibitem[{{Ishiguro} {et~al.}(2017){Ishiguro}, {Kuroda}, {Watanabe}, {Bach},
  {Kim}, {Lee}, {Sekiguchi}, {Naito}, {Ohtsuka}, {Hanayama}, {Hasegawa},
  {Usui}, {Urakawa}, {Imai}, {Sato}, \& {Kuramoto}}]{Ishiguro:2017}
{Ishiguro}, M., {Kuroda}, D., {Watanabe}, M., {et~al.} 2017, \aj, 154, 180

\bibitem[{Ishiguro {et~al.}(1997)Ishiguro, Nakayama, Kogachi, Mukai, Nakamura,
  Hirata, \& Okazaki}]{Ishiguro:1997}
Ishiguro, M., Nakayama, H., Kogachi, M., {et~al.} 1997, \pasj, 49, L31

\bibitem[{Ito {et~al.}(submitted)Ito, Arai, \& Ishiguro}]{Ito:2017}
Ito, T., Arai, T., \& Ishiguro, M. submitted

\bibitem[{{Jenniskens} {et~al.}(2009){Jenniskens}, {Shaddad}, {Numan}, {Elsir},
  {Kudoda}, {Zolensky}, {Le}, {Robinson}, {Friedrich}, {Rumble}, {Steele},
  {Chesley}, {Fitzsimmons}, {Duddy}, {Hsieh}, {Ramsay}, {Brown}, {Edwards},
  {Tagliaferri}, {Boslough}, {Spalding}, {Dantowitz}, {Kozubal}, {Pravec},
  {Borovicka}, {Charvat}, {Vaubaillon}, {Kuiper}, {Albers}, {Bishop},
  {Mancinelli}, {Sandford}, {Milam}, {Nuevo}, \& {Worden}}]{Jenniskens:2009}
{Jenniskens}, P., {Shaddad}, M.~H., {Numan}, D., {et~al.} 2009, \nat, 458, 485

\bibitem[{{Jenniskens} {et~al.}(2010){Jenniskens}, {Vaubaillon}, {Binzel},
  {DeMeo}, {Nesvorn{\'y}}, {Bottke}, {Fitzsimmons}, {Hiroi}, {Marchis},
  {Bishop}, {Vernazza}, {Zolensky}, {Herrin}, {Welten}, {Meier}, \&
  {Shaddad}}]{Jenniskens:2010}
{Jenniskens}, P., {Vaubaillon}, J., {Binzel}, R.~P., {et~al.} 2010, Meteoritics
  and Planetary Science, 45, 1590

\bibitem[{{Jewitt}(2012)}]{Jewitt:2012}
{Jewitt}, D. 2012, \aj, 143, 66

\bibitem[{Jockers {et~al.}(2005)Jockers, Kiselev, Bonev, Rosenbush,
  Shakhovskoy, Kolesnikov, Efimov, Shakhovskoy, \& Antonyuk}]{Jockers:2005}
Jockers, K., Kiselev, N., Bonev, T., {et~al.} 2005, \aap, 441, 773

\bibitem[{{Kaluna} {et~al.}(2017){Kaluna}, {Ishii}, {Bradley}, {Gillis-Davis},
  \& {Lucey}}]{Kaluna:2017}
{Kaluna}, H.~M., {Ishii}, H.~A., {Bradley}, J.~P., {Gillis-Davis}, J.~J., \&
  {Lucey}, P.~G. 2017, \icarus, 292, 245

\bibitem[{{Kim} {et~al.}(2014){Kim}, {Ishiguro}, \& {Usui}}]{Kim:2014}
{Kim}, Y., {Ishiguro}, M., \& {Usui}, F. 2014, \apj, 789, 151

\bibitem[{{Kiselev} {et~al.}(1990){Kiselev}, {Lupishko}, {Chernova}, \&
  {Shkuratov}}]{Kiselev:1990}
{Kiselev}, N.~N., {Lupishko}, D.~F., {Chernova}, G.~P., \& {Shkuratov}, I.~G.
  1990, Kinematika i Fizika Nebesnykh Tel, 6, 77

\bibitem[{Kiselev {et~al.}(1999)Kiselev, Rosenbush, \& Jockers}]{Kiselev:1999}
Kiselev, N.~N., Rosenbush, V.~K., \& Jockers, K. 1999, \icarus, 140, 464

\bibitem[{{Kiselev} {et~al.}(2002){Kiselev}, {Rosenbush}, {Jockers},
  {Velichko}, {Shakhovskoj}, {Efimov}, {Lupishko}, \&
  {Rumyantsev}}]{Kiselev:2002}
{Kiselev}, N.~N., {Rosenbush}, V.~K., {Jockers}, K., {et~al.} 2002, in ESA
  Special Publication, Vol. 500, Asteroids, Comets, and Meteors: ACM 2002, ed.
  B.~{Warmbein}, 887--890

\bibitem[{{Kohout} {et~al.}(2010){Kohout}, {Jenniskens}, {Shaddad}, \&
  {Haloda}}]{Kohout:2010}
{Kohout}, T., {Jenniskens}, P., {Shaddad}, M.~H., \& {Haloda}, J. 2010,
  Meteoritics and Planetary Science, 45, 1778

\bibitem[{Kuroda {et~al.}(2015)Kuroda, Ishiguro, Watanabe, Akitaya, Takahashi,
  Hasegawa, Ui, Kanda, Takaki, Itoh, Moritani, Imai, Goda, Takagi, Morihana,
  Honda, Arai, Hanayama, Nagayama, Nogami, Sarugaku, Murata, Morokuma, Saito,
  Oasa, Sekiguchi, \& Watanabe}]{Kuroda:2015}
Kuroda, D., Ishiguro, M., Watanabe, M., {et~al.} 2015, \apj, 814, 156

\bibitem[{Levasseur-Regourd {et~al.}(1996)Levasseur-Regourd, Hadamcik, \&
  Renard}]{Levasseur-Regourd:1996}
Levasseur-Regourd, A.~C., Hadamcik, E., \& Renard, J.~B. 1996, \aap, 313, 327

\bibitem[{{Lupishko}(2014)}]{Lupishko:2014}
{Lupishko}, D. 2014, NASA Planetary Data System, 215

\bibitem[{{Mainzer} {et~al.}(2014){Mainzer}, {Bauer}, {Cutri}, {Grav},
  {Masiero}, {Beck}, {Clarkson}, {Conrow}, {Dailey}, {Eisenhardt}, {Fabinsky},
  {Fajardo-Acosta}, {Fowler}, {Gelino}, {Grillmair}, {Heinrichsen}, {Kendall},
  {Kirkpatrick}, {Liu}, {Masci}, {McCallon}, {Nugent}, {Papin}, {Rice},
  {Royer}, {Ryan}, {Sevilla}, {Sonnett}, {Stevenson}, {Thompson}, {Wheelock},
  {Wiemer}, {Wittman}, {Wright}, \& {Yan}}]{Mainzer:2014}
{Mainzer}, A., {Bauer}, J., {Cutri}, R.~M., {et~al.} 2014, \apj, 792, 30

\bibitem[{{Mainzer} {et~al.}(2011){Mainzer}, {Grav}, {Bauer}, {Masiero},
  {McMillan}, {Cutri}, {Walker}, {Wright}, {Eisenhardt}, {Tholen}, {Spahr},
  {Jedicke}, {Denneau}, {DeBaun}, {Elsbury}, {Gautier}, {Gomillion}, {Hand},
  {Mo}, {Watkins}, {Wilkins}, {Bryngelson}, {Del Pino Molina}, {Desai},
  {G{\'o}mez Camus}, {Hidalgo}, {Konstantopoulos}, {Larsen}, {Maleszewski},
  {Malkan}, {Mauduit}, {Mullan}, {Olszewski}, {Pforr}, {Saro}, {Scotti}, \&
  {Wasserman}}]{Mainzer:2011}
{Mainzer}, A., {Grav}, T., {Bauer}, J., {et~al.} 2011, \apj, 743, 156

\bibitem[{{Masiero} \& {Cellino}(2009)}]{Masiero:2009}
{Masiero}, J. \& {Cellino}, A. 2009, \icarus, 199, 333

\bibitem[{Masiero {et~al.}(2012)Masiero, Mainzer, Grav, Bauer, Wright,
  McMillan, Tholen, \& Blain}]{Masiero:2012}
Masiero, J.~R., Mainzer, A.~K., Grav, T., {et~al.} 2012, \apj, 749, 104

\bibitem[{{Mieno} {et~al.}(2011){Mieno}, {Hasegawa}, \&
  {Mitsuishi}}]{Mieno:2011}
{Mieno}, T., {Hasegawa}, S., \& {Mitsuishi}, K. 2011, Japanese Journal of
  Applied Physics, 50, 125102

\bibitem[{{Moffat}(1969)}]{Moffat:1969}
{Moffat}, A.~F.~J. 1969, \aap, 3, 455

\bibitem[{{Mommert} {et~al.}(2014){Mommert}, {Hora}, {Harris}, {Reach},
  {Emery}, {Thomas}, {Mueller}, {Cruikshank}, {Trilling}, {Delbo}, \&
  {Smith}}]{Mommert:2014}
{Mommert}, M., {Hora}, J.~L., {Harris}, A.~W., {et~al.} 2014, \apj, 781, 25

\bibitem[{{Muinonen} {et~al.}(2002){Muinonen}, {Piironen}, {Shkuratov},
  {Ovcharenko}, \& {Clark}}]{Muinonen:2002}
{Muinonen}, K., {Piironen}, J., {Shkuratov}, Y.~G., {Ovcharenko}, A., \&
  {Clark}, B.~E. 2002, {Asteroid Photometric and Polarimetric Phase Effects}
  ({University of Arizona Press}), 123--138

\bibitem[{Nakamura {et~al.}(2011)Nakamura, Noguchi, Tanaka, Zolensky, Kimura,
  Tsuchiyama, Nakato, Ogami, Ishida, Uesugi, Yada, Shirai, Fujimura, Okazaki,
  Sandford, Ishibashi, Abe, Okada, Ueno, Mukai, Yoshikawa, \&
  Kawaguchi}]{Nakamura:2011}
Nakamura, T., Noguchi, T., Tanaka, M., {et~al.} 2011, \apss, 333, 1113

\bibitem[{{Nakayama} {et~al.}(2000){Nakayama}, {Fujii}, {Ishiguro}, {Nakamura},
  {Yokogawa}, {Yoshida}, \& {Mukai}}]{Nakayama:2000}
{Nakayama}, H., {Fujii}, Y., {Ishiguro}, M., {et~al.} 2000, \icarus, 146, 220

\bibitem[{{Noland} {et~al.}(1973){Noland}, {Veverka}, \&
  {Pollack}}]{Noland:1973}
{Noland}, M., {Veverka}, J., \& {Pollack}, J.~B. 1973, \icarus, 20, 490

\bibitem[{{Nugent} {et~al.}(2016){Nugent}, {Mainzer}, {Bauer}, {Cutri},
  {Kramer}, {Grav}, {Masiero}, {Sonnett}, \& {Wright}}]{Nugent:2016}
{Nugent}, C.~R., {Mainzer}, A., {Bauer}, J., {et~al.} 2016, \aj, 152, 63

\bibitem[{{Pearson} {et~al.}(2006){Pearson}, {Sephton}, {Franchi}, {Gibson}, \&
  {Gilmour}}]{Pearson:2006}
{Pearson}, V.~K., {Sephton}, M.~A., {Franchi}, I.~A., {Gibson}, J.~M., \&
  {Gilmour}, I. 2006, Meteoritics and Planetary Science, 41, 1899

\bibitem[{Penttil{\"a} {et~al.}(2005)Penttil{\"a}, Lumme, Hadamcik, \&
  Levasseur-Regourd}]{Penttila:2005}
Penttil{\"a}, A., Lumme, K., Hadamcik, E., \& Levasseur-Regourd, A.~C. 2005,
  \aap, 432, 1081

\bibitem[{{Pieters} \& {Hiroi}(2004)}]{Pieters:2004}
{Pieters}, C.~M. \& {Hiroi}, T. 2004, in Lunar and Planetary Science
  Conference, Vol.~35, Lunar and Planetary Science Conference, ed.
  S.~{Mackwell} \& E.~{Stansbery}

\bibitem[{{Popescu} {et~al.}(2012){Popescu}, {Birlan}, \&
  {Nedelcu}}]{Popescu:2012}
{Popescu}, M., {Birlan}, M., \& {Nedelcu}, D.~A. 2012, \aap, 544, A130

\bibitem[{{Sasaki} {et~al.}(2001){Sasaki}, {Nakamura}, {Hamabe}, {Kurahashi},
  \& {Hiroi}}]{Sasaki:2001}
{Sasaki}, S., {Nakamura}, K., {Hamabe}, Y., {Kurahashi}, E., \& {Hiroi}, T.
  2001, \nat, 410, 555

\bibitem[{{Schmidt} {et~al.}(1992){Schmidt}, {Elston}, \&
  {Lupie}}]{Schmidt:1992}
{Schmidt}, G.~D., {Elston}, R., \& {Lupie}, O.~L. 1992, \aj, 104, 1563

\bibitem[{{Shkuratov}(1985)}]{Shkuratov:1985}
{Shkuratov}, Y.~G. 1985, Astronomicheskij Tsirkulyar, 1400, 3

\bibitem[{Spencer {et~al.}(1989)Spencer, Lebofsky, \& Sykes}]{Spencer:1989}
Spencer, J.~R., Lebofsky, L.~A., \& Sykes, M.~V. 1989, \icarus, 78, 337

\bibitem[{{Tholen}(1984)}]{Tholen:1984}
{Tholen}, D.~J. 1984, PhD thesis, University of Arizona, Tucson

\bibitem[{{Tholen}(1989)}]{Tholen:1989}
{Tholen}, D.~J. 1989, {Asteroid taxonomic classifications} ({University of
  Arizona Press}), 1139--1150

\bibitem[{{Thomas} {et~al.}(1996){Thomas}, {Adinolfi}, {Helfenstein},
  {Simonelli}, \& {Veverka}}]{Thomas:1996}
{Thomas}, P.~C., {Adinolfi}, D., {Helfenstein}, P., {Simonelli}, D., \&
  {Veverka}, J. 1996, \icarus, 123, 536

\bibitem[{{Tinbergen}(1996)}]{Tinbergen:1996}
{Tinbergen}, J. 1996, {Astronomical Polarimetry} ({Cambridge University
  Press}), 174

\bibitem[{{Tody}(1993)}]{Tody:1993}
{Tody}, D. 1993, in Astronomical Society of the Pacific Conference Series,
  Vol.~52, Astronomical Data Analysis Software and Systems II, ed. R.~J.
  {Hanisch}, R.~J.~V. {Brissenden}, \& J.~{Barnes}, 173

\bibitem[{Tsuchiyama {et~al.}(2011)Tsuchiyama, Uesugi, Matsushima, Michikami,
  Kadono, Nakamura, Uesugi, Nakano, Sandford, Noguchi, Matsumoto, Matsuno,
  Nagano, Imai, Takeuchi, Suzuki, Ogami, Katagiri, Ebihara, Ireland, Kitajima,
  Nagao, Naraoka, Noguchi, Okazaki, Yurimoto, Zolensky, Mukai, Abe, Yada,
  Fujimura, Yoshikawa, \& Kawaguchi}]{Tsuchiyama:2011}
Tsuchiyama, A., Uesugi, M., Matsushima, T., {et~al.} 2011, \apss, 333, 1125

\bibitem[{Umow(1905)}]{Umow:1905}
Umow, N. 1905, Physikalische Zeitschrift, 6, 674

\bibitem[{{Vernazza} {et~al.}(2017){Vernazza}, {Castillo-Rogez}, {Beck},
  {Emery}, {Brunetto}, {Delbo}, {Marsset}, {Marchis}, {Groussin}, {Zanda},
  {Lamy}, {Jorda}, {Mousis}, {Delsanti}, {Djouadi}, {Dionnet}, {Borondics}, \&
  {Carry}}]{Vernazza:2017}
{Vernazza}, P., {Castillo-Rogez}, J., {Beck}, P., {et~al.} 2017, \aj, 153, 72

\bibitem[{{Vernazza} {et~al.}(2015){Vernazza}, {Marsset}, {Beck}, {Binzel},
  {Birlan}, {Brunetto}, {Demeo}, {Djouadi}, {Dumas}, {Merouane}, {Mousis}, \&
  {Zanda}}]{Vernazza:2015}
{Vernazza}, P., {Marsset}, M., {Beck}, P., {et~al.} 2015, \apj, 806, 204

\bibitem[{{Warner}(2016)}]{Warner:2016}
{Warner}, B.~D. 2016, Minor Planet Bulletin, 43, 143

\bibitem[{{Watanabe} {et~al.}(2012){Watanabe}, {Takahashi}, {Sato}, {Watanabe},
  {Fukuhara}, {Hamamoto}, \& {Ozaki}}]{Watanabe:2012}
{Watanabe}, M., {Takahashi}, Y., {Sato}, M., {et~al.} 2012, in \procspie, Vol.
  8446, Ground-based and Airborne Instrumentation for Astronomy IV, 84462O

\bibitem[{{Whiteley}(2001)}]{Whiteley:2001}
{Whiteley}, Jr., R.~J. 2001, PhD thesis, University of Hawai'i at Manoa

\bibitem[{{Worms} {et~al.}(2000){Worms}, {Renard}, {Hadamcik}, {Brun-Huret}, \&
  {Levasseur-Regourd}}]{Worms:2000}
{Worms}, J.-C., {Renard}, J.-B., {Hadamcik}, E., {Brun-Huret}, N., \&
  {Levasseur-Regourd}, A.~C. 2000, \planss, 48, 493

\bibitem[{{Zellner} {et~al.}(1974){Zellner}, {Gehrels}, \&
  {Gradie}}]{Zellner:1974a}
{Zellner}, B., {Gehrels}, T., \& {Gradie}, J. 1974, \aj, 79, 1100

\bibitem[{Zellner \& Gradie(1976)}]{Zellner:1976}
Zellner, B. \& Gradie, J. 1976, \aj, 81, 262

\bibitem[{{Zellner} \& {Capen}(1974)}]{Zellner:1974b}
{Zellner}, B.~H. \& {Capen}, R.~C. 1974, \icarus, 23, 437

\end{thebibliography}

\end{document}